\documentclass[10pt, conference, letterpaper]{IEEEtran}
\IEEEoverridecommandlockouts
\usepackage{float}  
\usepackage{stfloats}
\usepackage{cite}
\usepackage{amsmath,amssymb,amsfonts}
\usepackage{amsmath}  
\usepackage{subcaption}
\usepackage{graphicx}
\usepackage{booktabs}
\usepackage{textcomp}
\usepackage{multirow}
\usepackage{xcolor}
\usepackage{xspace}
\usepackage{algorithm}
\usepackage{algpseudocode}
\usepackage{tabularx, makecell}
\usepackage{makecell}
\usepackage{caption} 
\usepackage{listings}
\def\BibTeX{{\rm B\kern-.05em{\sc i\kern-.025em b}\kern-.08em
    T\kern-.1667em\lower.7ex\hbox{E}\kern-.125emX}}
\begin{document}

\title{\vspace{-1em} FLEET: A Federated Learning Emulation and Evaluation Testbed for Holistic Research \\
\vspace{0.2em}
\footnotesize This work has been submitted to the IEEE for possible publication.\\ Copyright may be transferred without notice, after which this version may no longer be accessible.
}

\author{
\IEEEauthorblockN{Osama Abu Hamdan, Hao Che}
\IEEEauthorblockA{
\textit{University of Texas at Arlington}\\
oma8085@mavs.uta.edu, hche@cse.uta.edu}
\and
\IEEEauthorblockN{Engin Arslan}
\IEEEauthorblockA{\textit{Meta} \\
enginarslan@meta.com}
\and
\IEEEauthorblockN{Md Arifuzzaman}
\IEEEauthorblockA{
\textit{Missouri University of Science and Technology}\\
marifuzzaman@mst.edu}
}

\maketitle

\newcommand{\name}{\textit{FLEET}\xspace}

\begin{abstract}
Federated Learning (FL) presents a robust paradigm for privacy-preserving, decentralized machine learning. However, a significant gap persists between the theoretical design of FL algorithms and their practical performance, largely because existing evaluation tools often fail to model realistic operational conditions. Many testbeds oversimplify the critical dynamics among algorithmic efficiency, client-level heterogeneity, and continuously evolving network infrastructure. To address this challenge, we introduce the Federated Learning Emulation and Evaluation Testbed (\name). This comprehensive platform provides a scalable and configurable environment by integrating a versatile, framework-agnostic learning component with a high-fidelity network emulator. \name supports diverse machine learning frameworks, customizable real-world network topologies, and dynamic background traffic generation. The testbed collects holistic metrics that correlate algorithmic outcomes with detailed network statistics. By unifying the entire experiment configuration, \name enables researchers to systematically investigate how network constraints, such as limited bandwidth, high latency, and packet loss, affect the convergence and efficiency of FL algorithms. This work provides the research community with a robust tool to bridge the gap between algorithmic theory and real-world network conditions, promoting the holistic and reproducible evaluation of federated learning systems.
\end{abstract}


\section{Introduction}
Federated Learning (FL) has emerged as a transformative approach to distributed machine learning, enabling collaborative model training on decentralized devices while ensuring user data remains localized \cite{mcmahan2017communication}. This paradigm is crucial for domains such as healthcare, the Internet of Things (IoT), and autonomous systems, where centralizing data is often infeasible due to privacy regulations, data sovereignty concerns, or prohibitive communication costs. By orchestrating model training directly at the network edge, FL leverages vast distributed datasets to build powerful models without compromising user privacy, signaling a fundamental departure from traditional, centralized artificial intelligence.

Despite its promise, the practical deployment of robust and efficient FL systems faces significant hurdles, particularly in the domain of evaluation. The performance of an FL algorithm is not solely a product of its mathematical design; it is deeply coupled with the environment in which it operates. This environment is defined by a triad of challenges: statistical heterogeneity, where data is non-independently and non-identically (non-IID) distributed across clients; systems heterogeneity, where devices possess varied computational, memory, and power resources; and network dynamics, where communication links suffer from fluctuating bandwidth, latency, and reliability~\cite{li2020federated}. While benchmarks like LEAF~\cite{caldas2019leaf} have made important advances in providing realistic federated datasets to address statistical heterogeneity~\cite{caldas2019leaf}, a truly comprehensive evaluation must also integrate the complexities of the underlying system and networking infrastructure.

To properly assess FL algorithms, researchers require tools that can comprehensively model these multifaceted challenges. Current evaluation methodologies, however, reveal a distinct trade-off between realism and scale. Pure software simulations offer scalability but typically rely on abstract models of network and system behavior, which may fail to capture the complex, non-linear interactions that arise in real-world deployments. This abstraction can create a significant "simulation-to-reality" gap, where algorithms that appear effective in theory underperform in practice. Conversely, physical hardware testbeds provide high-fidelity performance data but are expensive to build, difficult to maintain, and lack the flexibility to systematically study diverse network topologies or dynamic conditions at scale. There is a pressing need for a platform that merges the configurability of simulation with the realism of hardware environments.

In response to these challenges, we have developed the Federated Learning Emulation and Evaluation Testbed (\textit{FLEET}). Our work directly addresses the need for high-fidelity evaluation by innovating a platform around the synergy between the Flower AI framework~\cite{beutel2020flower} and the Containernet network emulator~\cite{peuster2016medicine}. This integration yields a powerful and flexible testing environment. Flower’s framework-agnostic nature grants researchers the freedom to experiment with familiar tools like PyTorch, TensorFlow, and JAX. Simultaneously, Containernet leverages Docker containers to create realistic client emulations with fine-grained control over system resources and network link characteristics. This architecture allows for the recreation of complex, real-world network topologies and the simulation of dynamic conditions through background traffic injection. By collecting and synchronizing comprehensive metrics that span both algorithmic performance and network statistics, \textit{FLEET} provides a holistic view of an FL system in operation, thereby enabling a more rigorous and insightful investigation into the multifaceted challenges of federated learning. \name is available at \cite{code}.

\section{Related Work}
The evaluation of FL systems has inspired various research tools and testbeds, each addressing different challenges of decentralized training. Our work builds upon the foundations laid by key testbeds that have progressively introduced more practical consideration into FL experimentation. We position our contribution concerning three significant works, LEAF, CoLexT \cite{bozic2024where}, and MininetFed \cite{sarmento2024mininetfed}, to highlight the specific gaps our integrated platform, \name, aims to fill.

\begin{table*}[b]
  \centering
  \caption{Comprehensive Comparison of Federated Learning Testbeds Highlighting Their Core Paradigms, System Architectures, Network Emulation Capabilities, Configurability, and Performance Metrics}
  \label{tab:related_comparison}
  \renewcommand{\arraystretch}{1.0} 
  \begin{tabular}{@{} l l l l l @{}}
    \toprule
    \textbf{Feature} & \textbf{LEAF} & \textbf{CoLexT} & \textbf{MininetFed} & \name \\
    \midrule
    \textbf{Core Paradigm} & 
      \makecell[l]{Statistical heterogeneity \\ benchmark} & 
      \makecell[l]{Real-world hardware \\ testbed} & 
      \makecell[l]{System emulation} & 
      \makecell[l]{System emulation} \\
    \midrule
    \textbf{FL Framework} & 
      \makecell[l]{Custom reference \\ implementations} & 
      \makecell[l]{Flower \\ (Framework-Agnostic)} & 
      \makecell[l]{Custom implementation \\ (MQTT-based)} & 
      \makecell[l]{Flower \\ (Framework-Agnostic)} \\
    \midrule
    \textbf{Topology Support} & 
      Not applicable & 
      \makecell[l]{Fixed \\ (Physical network)} & 
      Custom scripts & 
      \makecell[l]{Extensive: (Topohub, \\ Internet Topology Zoo, \\ custom scripts)} \\
    \midrule
    \textbf{Background Traffic} & 
      No & 
      No & 
      No & 
      \makecell[l]{Iperf3 or TCPReplay with \\ multiple traffic patterns \\} \\
    \midrule
    \textbf{Key Metrics} & 
      \makecell[l]{FL (accuracy), \\ System (FLOPS, bytes)} & 
      \makecell[l]{FL (accuracy), \\ System (CPU, energy, \\ memory)} & 
      \makecell[l]{FL (accuracy, loss), \\ Network (traffic)} & 
      \makecell[l]{FL (accuracy, loss), \\ Network (TX/RX), \\ System (CPU, memory)} \\
    \midrule
    \textbf{Primary Contribution} & 
      \makecell[l]{Standardized datasets and \\ metrics for statistical \\ heterogeneity.} & 
      \makecell[l]{High-fidelity system metrics \\ (especially energy) from \\ physical hardware.} & 
      \makecell[l]{Flexible emulation of client \\ resources and network links \\ with security.} & 
      \makecell[l]{Unified, realistic emulation \\ of FL algorithms, systems, \\ and networks.} \\
    \bottomrule
  \end{tabular}
\end{table*} 

Caldas et al. introduced LEAF, a foundational benchmark designed to address statistical heterogeneity in federated settings. Its primary contribution is a suite of open-source datasets with natural user-based partitions that model non-IID and unbalanced data distributions. LEAF provides a common ground for comparing algorithms based on how they handle data challenges, focusing primarily on the data layer of FL evaluation.

To incorporate systems-level realism, Božič et al. developed CoLexT, a testbed for FL experimentation on physical devices. CoLexT excels at capturing authentic hardware performance metrics, such as CPU utilization and real-world energy usage, by deploying algorithms directly onto a heterogeneous collection of single-board computers and smartphones. This approach provides high-fidelity system measurements that are unattainable in pure simulation.

Sarmento et al. proposed MininetFed, a container-based emulation tool for modeling FL environments. It allows researchers to configure heterogeneous client hardware and network link properties, such as bandwidth and latency, within a containerized architecture. This provides a flexible platform for exploring the interplay between FL algorithms and the system and network parameters of the clients.

Our testbed, \name, synthesizes the strengths of these prior works while addressing their individual limitations. It extends beyond the purely statistical focus of LEAF by providing a rich environment to emulate critical network and systems-level challenges. Unlike CoLexT, which relies on physical hardware, the emulation-based approach of \name offers superior scalability, cost-effectiveness, and the flexibility to model thousands of complex network topologies. Finally, \name improves upon MininetFed by integrating with the widely-adopted Flower ecosystem for broad compatibility and by incorporating more advanced network modeling. Specifically, it utilizes real-life network topologies from sources like Topohub~\cite{topohub} and introduces background traffic with different patterns, capabilities not extensively supported in MininetFed. This holistic approach provides a unified platform for comprehensive FL research. Table \ref{tab:related_comparison} provides a detailed comparison of the core paradigms, architectures, and capabilities across these testbeds and our own.

\section{System Overview}

\begin{figure*}[t]
  \centering
  \includegraphics[width=\textwidth]{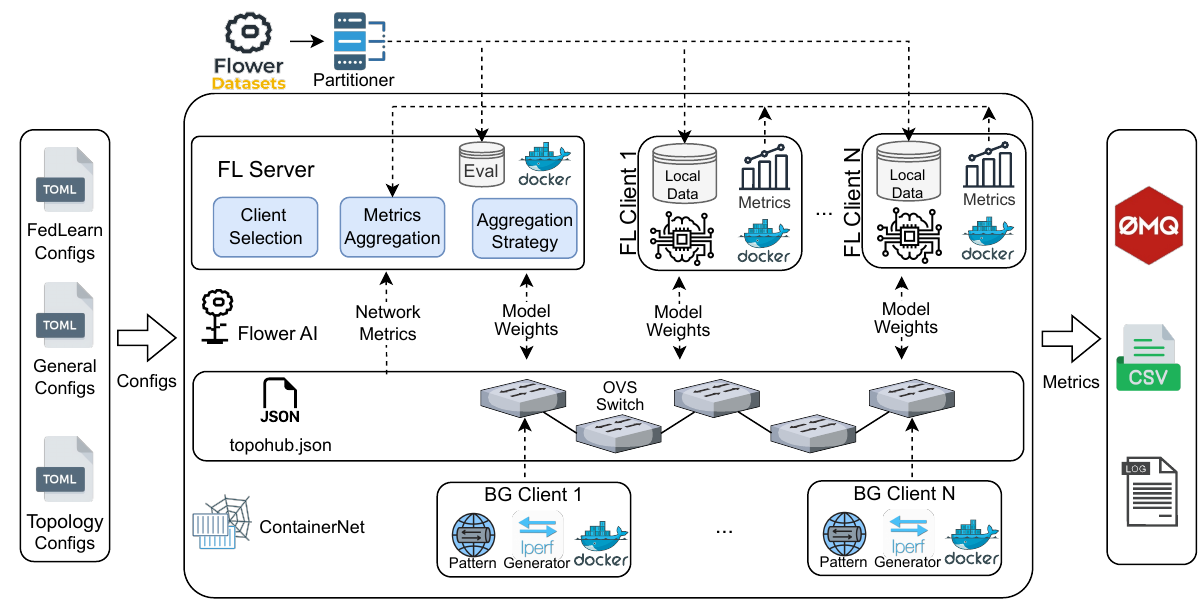}
  \caption{System architecture of \name, integrating Flower for Federated Learning and Containernet for network emulation. The system is configured via TOML files, uses Topohub for network topologies, and generates background traffic to simulate realistic conditions. Comprehensive metrics from all components are exported to ZMQ, CSV, or log files for analysis.}
  \label{fig:your_label}
\end{figure*}

We designed a highly configurable and extensible testbed that integrates a robust FL framework with a realistic network emulation environment. The core architecture combines \textbf{Flower}, a framework-agnostic FL library, with \textbf{Containernet}, a Docker-based network emulator. This integration enables researchers to conduct reproducible experiments that jointly investigate algorithmic performance and network dynamics. Each FL participant operates within an individual Docker container, connected through an emulated network where link characteristics are precisely controlled.

The entire experimental workflow is orchestrated through a unified configuration system using a set of three TOML files for FL, Containernet, and general settings. This separation promotes ease of use and reproducibility, and we provide default settings for a quick start. A user defines all experimental parameters across these files, from which the testbed automatically sets up the network, deploys the containers, initiates training, generates background traffic, and collects comprehensive metrics. This holistic approach bridges the gap between purely theoretical FL simulations and complex, real-world deployments.

\subsection{Federated Learning Component}
The testbed leverages \textit{Flower} framework, a flexible and extensible platform for building federated learning systems. Flower supports a wide range of machine learning models, aggregation strategies, and deployment scenarios, making it well-suited for research and experimentation in distributed learning. It provides built-in abstractions for client-server communication, training orchestration, and evaluation, allowing researchers to prototype and scale federated learning workflows with minimal boilerplate.

To better support our experimental goals, we extended Flower’s server and client components to enable advanced metrics collection and greater configurability. Users can easily customize or replace core components, such as models, datasets, and aggregation methods, via configuration files. Additionally, we integrated the \texttt{psutil} library to collect system-level resource usage from clients, enabling the server to implement resource-aware client selection strategies. These capabilities are not natively supported by Flower and were essential for our research focus on performance and resource-constrained environments.

\subsection{Dataset Handling and Partitioning}
A critical aspect of realistic FL emulation is the ability to model heterogeneous data distributions across clients. Our testbed incorporates the \textit{Flower Datasets} library, which provides a powerful abstraction for dataset access and partitioning. This library integrates seamlessly with the HuggingFace Hub, granting access to thousands of datasets across various domains, such as standard image classification benchmarks (\texttt{CIFAR-10}, \texttt{MNIST}) and natural language processing tasks (\texttt{IMDB}, \texttt{WikiText}).

To simulate data heterogeneity, the testbed supports multiple partitioning strategies. A common baseline is \texttt{IID partitioning}, where data is distributed uniformly among clients. To model more realistic scenarios, it also provides several \texttt{non-IID mechanisms}. For instance, a shard-based approach divides the dataset by class label, assigning a limited number of classes to each client. A more fine-grained method uses a \texttt{Dirichlet distribution} to vary class concentrations across clients, allowing for precise control over the degree of statistical heterogeneity.

\subsection{Network Emulation with Containernet}
We use \textit{Containernet} to construct the underlying network fabric for our experiments. As a fork of the Mininet emulator, Containernet replaces standard hosts with Docker containers, enabling full application stack execution within each network node. This container-based approach provides process and filesystem isolation for each FL client, closely mimicking a distributed production environment.

The network topology itself is highly customizable. A key feature is the integration with \textit{Topohub}, an online repository that provides thousands of real-world and synthetic network models. This allows experiments to be grounded in realistic network structures, a critical factor for validating results. Alternatively, users can define custom structures (e.g., star, mesh) via a standard \texttt{Mininet Python script}. Beyond topology, the testbed offers granular control over link characteristics like bandwidth, latency, and loss rate, as well as per-container resource limits to simulate heterogeneous client hardware.

\subsection{Background Traffic Generation}
In real-world networks, FL communication competes for bandwidth with other applications. To simulate this crucial aspect, the testbed includes a modular component for generating background traffic. This component deploys dedicated Docker containers that act as traffic generators and sinks, creating network congestion to test the resilience of FL algorithms under realistic load.

The module provides built-in support for two widely-used generation tools: \texttt{Iperf3} for creating TCP and UDP flows, and \texttt{TCPReplay} for replaying captured network traffic. The intensity and timing of this traffic can follow various statistical patterns, summarized in Table~\ref{tab:traffic_patterns}, to simulate different network scenarios. The architecture of this component is intentionally modular, so advanced users can easily integrate their own traffic generation tools or define custom patterns.

\begin{table}[b]
\centering
\caption{Background Traffic Patterns Supported by \name}
\label{tab:traffic_patterns}
\renewcommand{\arraystretch}{1.3}
\begin{tabular}{@{}ll@{}}
\toprule
\textbf{Traffic Pattern} & \textbf{Description} \\ \midrule
\textit{Poisson} & Simulates random, memoryless event arrivals. \\
\textit{Bursty} & Alternates between high-traffic bursts and idle periods. \\
\textit{Uniform} & Maintains a constant traffic rate over time. \\
\textit{Normal} & Varies traffic rate according to a Gaussian distribution. \\
\textit{SineWave} & Generates traffic with smooth, sinusoidal variation. \\
\bottomrule
\end{tabular}
\end{table}

\subsection{Metrics Aggregation and Logging}
A comprehensive evaluation requires correlating application-level performance with underlying system behavior. Our testbed is designed for holistic data collection, capturing metrics from all system components. The metrics fall into two main categories:
\begin{enumerate}
    \item \textbf{Federated Learning Metrics}: These include model performance indicators like training loss and evaluation accuracy, round timings, and derived values such as communication time.
    \item \textbf{System and Network Metrics}: These include the extended, client-specific metrics such as CPU and memory utilization, alongside network interface statistics, such as TX/RX bytes and rates, collected from Containernet.
\end{enumerate}

To accommodate different analysis workflows, the testbed provides multiple logging destinations. For offline analysis, all metrics can be saved to local Logfiles and CSV files, which are easily imported into standard data analysis tools. For real-time monitoring, the testbed can stream all metrics over a \texttt{ZeroMQ (ZMQ)} publisher-subscriber socket, enabling the connection of external visualization dashboards. This multi-faceted approach ensures that researchers can capture the rich data needed to fully understand the complex interactions within the system.

\section{Experimental Results}\label{sec:results}
To validate the capabilities and demonstrate the versatility of our proposed testbed, we conducted a series of experiments designed to showcase its utility for comprehensive FL research, spanning real-time performance monitoring, in-depth analysis of network dynamics, comparative studies of learning algorithms, and the impact of realistic network congestion. The testbed has also been successfully used in other research work, including SmartFLow \cite{smartflow}, demonstrating its practical applicability. The results underscore its ability to facilitate reproducible research into the complex interplay between FL algorithms, system resources, and network conditions.

\subsection{Real-time Monitoring and Visualization}
A key feature of our testbed is its capacity for real-time experiment monitoring, enabled by the integration of a ZeroMQ (ZMQ) publisher. This component streams a wide array of metrics from the server and all clients to any subscribed listener. We developed a Plotly-based dashboard to connect to this ZMQ stream, providing live visualization of the training process. This immediate feedback is invaluable for debugging, observing convergence patterns, and understanding system behavior without waiting for post-hoc analysis.

\begin{figure*}[htbp]
    \centering
    \begin{subfigure}[t]{0.49\textwidth}
        \centering
        \includegraphics[width=\linewidth]{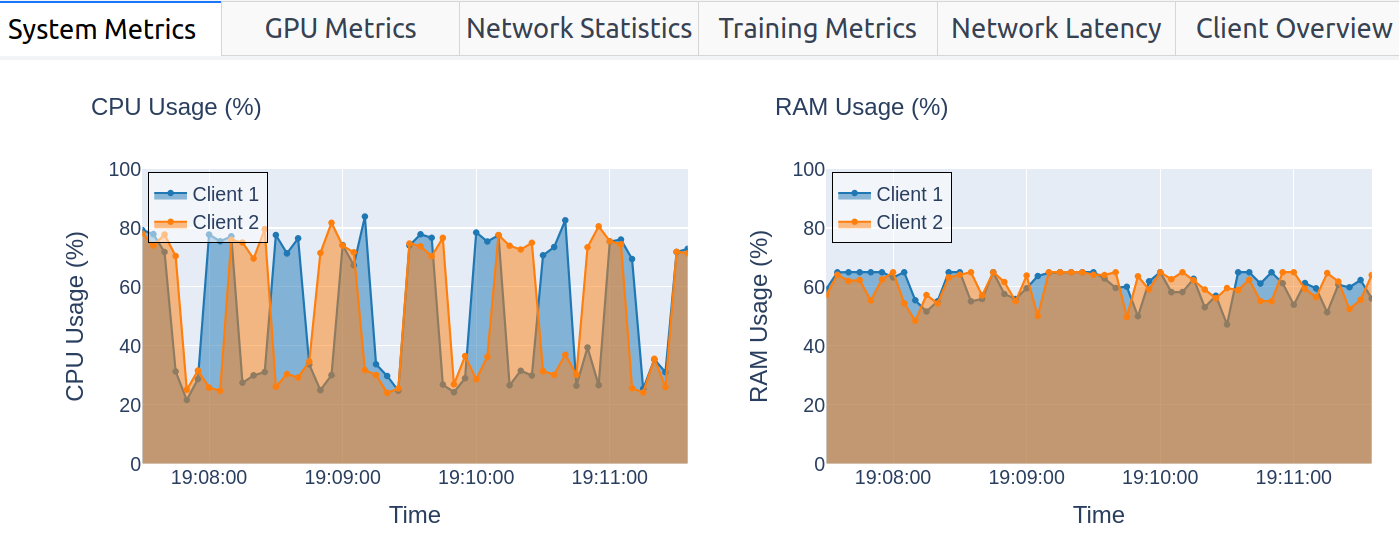}
        \caption{Per-client system resources (CPU and RAM) during training.}
        \label{fig:fig1}
    \end{subfigure}
    \begin{subfigure}[t]{0.49\textwidth}
        \centering
        \includegraphics[width=\linewidth]{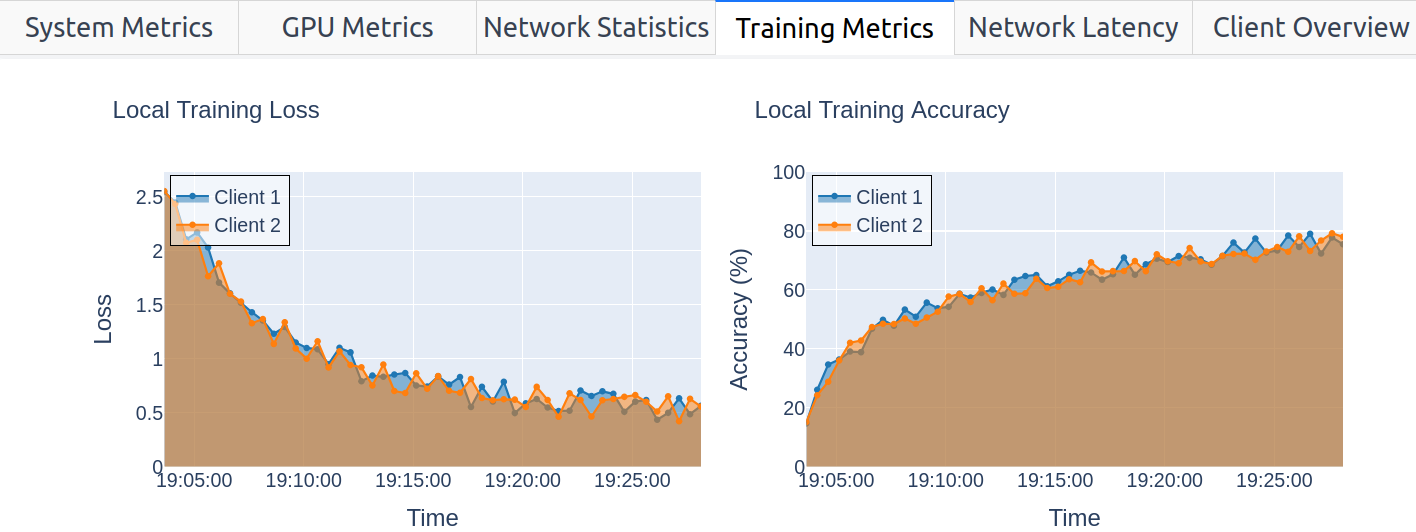}
        \caption{Per-client FL metrics (loss and accuracy) during training.}
        \label{fig:fig2}
    \end{subfigure}
    \caption{Real-time system and FL metrics streamed via ZMQ for real-time monitoring.}
    \label{fig:full_monitoring}
    \vspace{-2mm}
\end{figure*}

Figure~\ref{fig:fig1} and Figure~\ref{fig:fig2} illustrate the dashboard's capabilities. Figure~\ref{fig:fig1} displays the per-client CPU and memory utilization over time. This view allows a researcher to directly observe the computational footprint of the FL task on each device, providing critical insights into resource-constrained clients and the potential for system-aware client selection strategies. Simultaneously, Figure~\ref{fig:fig2} plots the progression of training loss and evaluation accuracy for each participating client. The ability to observe both system and learning metrics provides a holistic view of the experiment as it unfolds.

\subsection{Comprehensive Metrics for Holistic Analysis}
Beyond real-time visualization, the testbed logs extensive metrics to CSV files, enabling detailed offline analysis. This rich dataset allows researchers to investigate phenomena that are not apparent from high-level performance indicators alone. We demonstrate this through the collection of granular network statistics and by conducting comparative algorithmic studies under diverse data distributions.

\subsubsection{Network Performance Analysis}
\begin{figure}[t]
    \centering
    \begin{subfigure}[t]{0.49\linewidth}
        \centering
        \includegraphics[width=\linewidth]{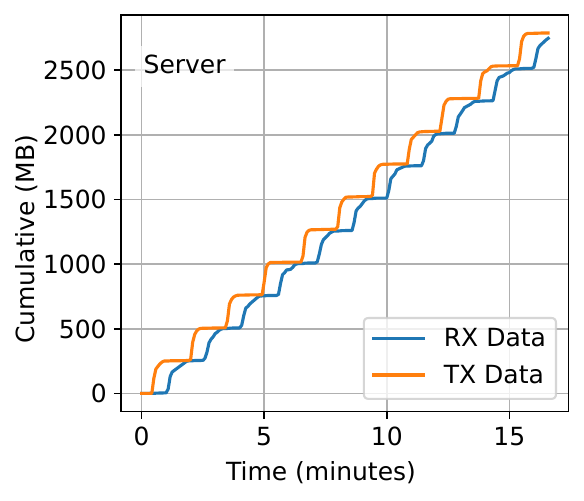}
        \caption{FL server TX/RX bytes.}
        \label{fig:fig3}
    \end{subfigure}
    \begin{subfigure}[t]{0.49\linewidth}
        \centering
        \includegraphics[width=\linewidth]{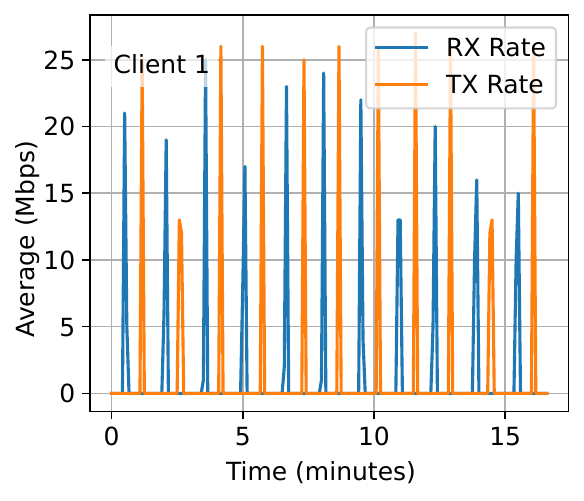}
        \caption{FL client TX/RX rates.}
        \label{fig:fig4}
    \end{subfigure}
    \caption{Network traffic rates (TX/RX bps) and cumualtive data (TX/RX MB) during training, collected via custom script and logged to a CSV file.}
    \label{fig:network_monitoring}
    \vspace{-2mm}
\end{figure}

Our testbed employs custom scripts to capture detailed network interface statistics from each container. Figure~\ref{fig:fig3} shows the cumulative transmitted (TX) and received (RX) data for the central FL server, while Figure~\ref{fig:fig4} presents the TX and RX rate rates for a single FL client. Such data allows researchers to identify communication bottlenecks, correlate network activity with specific FL events, and analyze the fine-grained network behavior of different aggregation protocols.

\subsubsection{Algorithmic and Data Heterogeneity}
The true power of an FL testbed lies in its ability to systematically explore the vast parameter space of federated learning. We showcase this by conducting an experiment using the CIFAR-10 dataset~\cite{krizhevsky2009learning} with a MobileNetV3-Large model~\cite{howard2019searching}, deployed across 15 clients in a network topology from the Gabriel collection in Topohub. We evaluated three distinct data partitioning schemes and their interaction with different aggregation algorithms. Specifically, we used Pathological non-IID with FedYogi, Dirichlet-based non-IID with FedAvgM, and an IID setup with FedProx. 

\begin{figure}[t]
    \centering
    \begin{subfigure}[t]{0.49\linewidth}
        \centering
        \includegraphics[width=\linewidth]{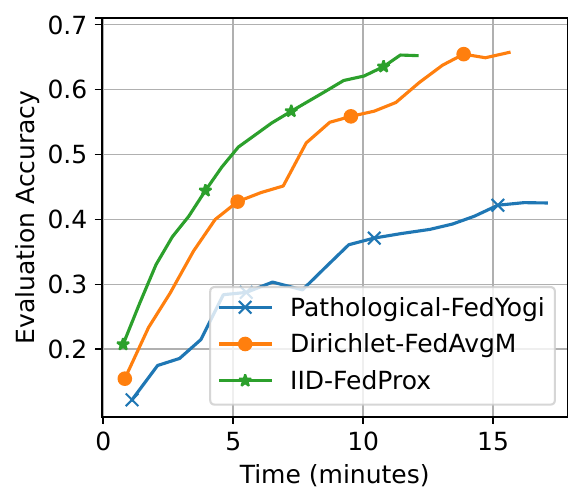}
        \caption{Evaluation Accuracy}
        \label{fig:mp_acc}
    \end{subfigure}
    \begin{subfigure}[t]{0.49\linewidth}
        \centering
        \includegraphics[width=\linewidth]{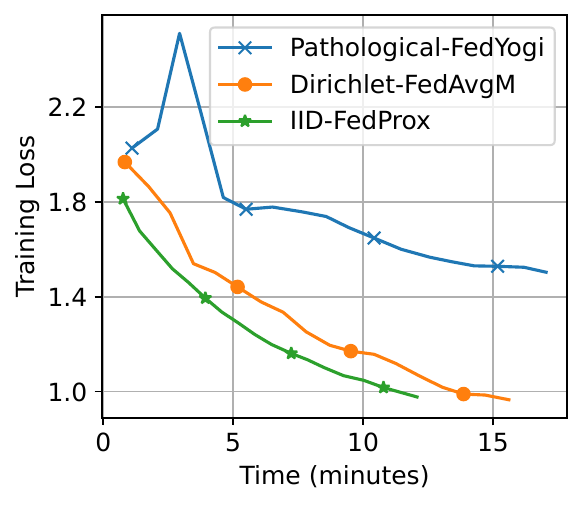}
        \caption{Training Loss}
        \label{fig:mp_loss}
    \end{subfigure}
    \caption{Global model accuracy and loss under varied data partitions (Pathological, Dirichlet, IID) and aggregation algorithms (FedYogi, FedAvgM, FedProx).}
    \label{fig:mp}
    \vspace{-2mm}
\end{figure}

The results in Figure~\ref{fig:mp} demonstrate the testbed's ability to produce nuanced, comparative insights. \figurename~\ref{fig:mp_acc} presents the evaluation accuracy curves, while \figurename~\ref{fig:mp_loss} shows the corresponding loss trajectories. Together, these plots enable researchers to analyze not only the final performance metrics, but also the convergence behavior of different method combinations. For instance, a researcher can use such results to quantify the performance degradation an algorithm suffers when moving from IID to non-IID data, or to identify which algorithm is most robust to a specific type of data heterogeneity.

\subsection{Emulating Realistic Network Conditions}
Many FL simulation tools operate under an unrealistic assumption of ideal network conditions, which limits their applicability. Our testbed addresses this limitation by integrating a robust module for generating background network traffic. We demonstrate its impact through an experiment using the IMDB movie review dataset~\cite{maas-EtAl:2011:ACL-HLT2011} and a DistilBERT model~\cite{sanh2019distilbert}. The setup involved 10 FL clients connected via the Nobel-EU topology, a real-world research network model from Topohub.

\begin{figure*}[htbp]
\centering
\begin{minipage}[t]{0.49\textwidth}
    \centering
    \begin{subfigure}[t]{0.49\textwidth}
        \centering
        \includegraphics[width=\linewidth]{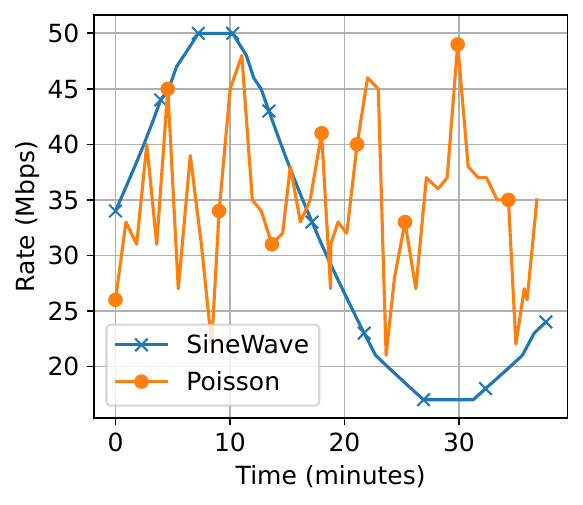}
        \caption{Variant BG traffic patterns}
        \label{fig:fig8}
    \end{subfigure}
    \begin{subfigure}[t]{0.49\textwidth}
        \centering
        \includegraphics[width=\linewidth]{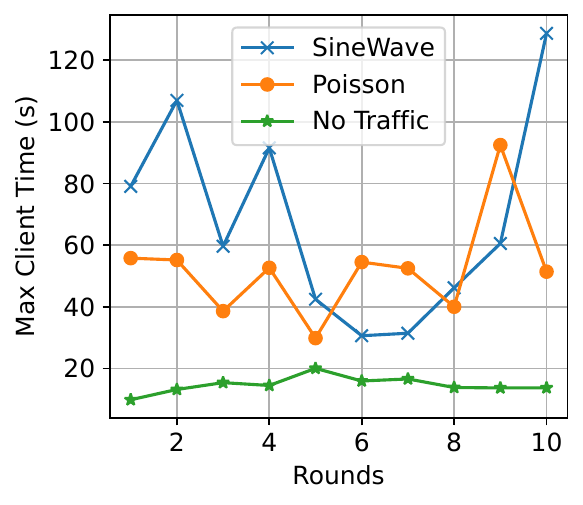}
        \caption{S2C max client latency}
        \label{fig:fig7}
    \end{subfigure}%
    \caption{Impact of background traffic on federated learning communication. Both Poisson and SineWave patterns increase latency for the slowest client compared to a no-traffic baseline, with the Poisson pattern exhibiting higher variability.}
    \label{fig:main_right}
\end{minipage}
\hfill
\begin{minipage}[t]{0.49\textwidth}
    \centering
    \begin{subfigure}[t]{0.49\textwidth}
        \centering
        \includegraphics[width=\linewidth]{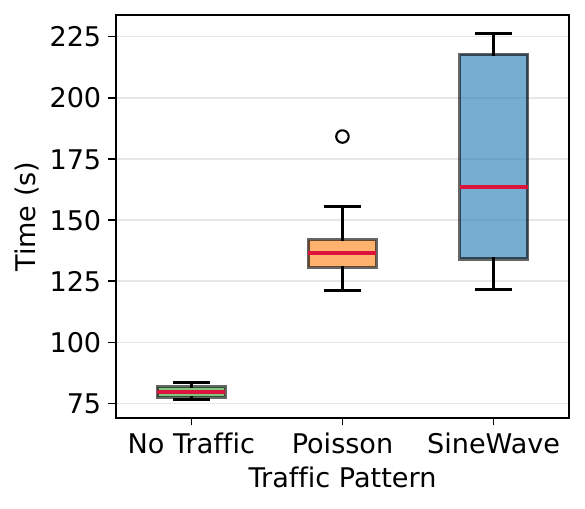}
        \caption{Round completion time distribution}
        \label{fig:fig6}
    \end{subfigure}
    \begin{subfigure}[t]{0.49\textwidth}
        \centering
        \includegraphics[width=\linewidth]{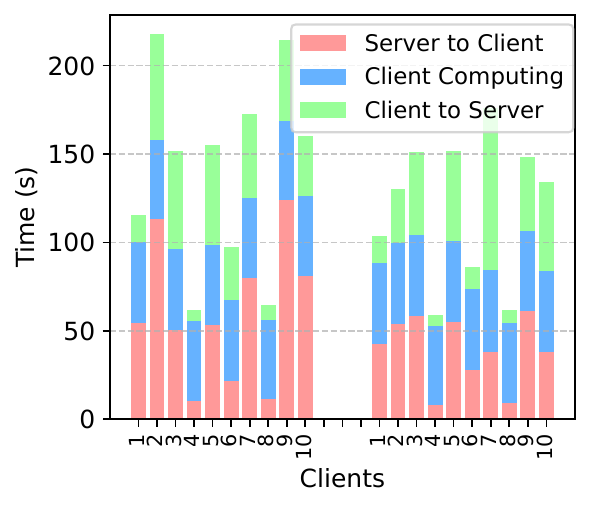}
        \caption{Breakdown of round time components}
        \label{fig:fig5}
    \end{subfigure}
    \caption{Analysis of federated learning round times under various network conditions. The distribution of overall round times reveals the impact of traffic patterns, while the component breakdown identifies specific bottlenecks.}
    \label{fig:main_left}
\end{minipage}
\end{figure*}

We used Iperf3 to generate background traffic between nodes, creating congestion that competes with FL communication. Figure~\ref{fig:fig8} illustrates two of the supported traffic patterns: a smooth, predictable \textit{SineWave} pattern and a bursty, unpredictable \textit{Poisson} pattern. Both patterns were capped at a bandwidth of 50 Mbps on a specific link. The background traffic profoundly affects FL performance. As Figure~\ref{fig:fig7} shows, introducing traffic significantly increases communication time compared to a no-traffic baseline. The nature of the traffic also matters, with the bursty Poisson pattern inducing higher and more variable latency than the smoother SineWave pattern. This result highlights the critical need for realistically modeling network dynamics, as an FL algorithm's performance can depend heavily on the characteristics of the underlying network traffic.

To analyze these effects in more detail, we dissected the time spent in a single FL round. Figure~\ref{fig:fig6} compares the distribution of round completion times across different traffic patterns, illustrating how the overall round duration varies. For a more granular view, Figure~\ref{fig:fig5} breaks down the time each client spends in distinct phases: S2C communication , local computation, C2S communication. This detailed timing information, collected via CSV logging, enables researchers to precisely identify bottlenecks. For example, a researcher can determine whether a long round time is due to a slow network link (high S2C/C2S time) or an underpowered client (high computation time). This granular analysis is essential for interpreting the effects of network congestion and for developing network-aware scheduling or aggregation strategies that are resilient to real-world conditions.

\section{Conclusion and Future Work}\label{sec:conlusion}
We introduced \name, a comprehensive and extensible federated learning testbed designed to bridge the gap between abstract simulation and real-world network conditions. By integrating the Flower federated learning framework with the Containernet network emulator, our system provides a high-fidelity environment for evaluating the performance and resilience of FL algorithms. The architecture allows researchers to control experimental variables with precision, from data partitioning and model selection to network topology and background traffic. Key contributions of this work include the modular configuration system using TOML files, the integration with Topohub for realistic network topologies, and the extension of the Flower framework to capture client-side system metrics. This holistic approach enables reproducible research that jointly considers algorithmic behavior and the impact of underlying network and hardware heterogeneity, providing a critical tool for developing robust and efficient federated systems.

While this work provides a robust foundation, we have identified several promising directions for future development. A primary goal is to extend the emulation capabilities to include {wireless networks, particularly Wi-Fi and cellular standards, to model the complexities of signal attenuation, interference, and client mobility. To support larger-scale experiments, we will evolve the testbed to operate in a distributed fashion across multiple physical hosts, leveraging graph partitioning libraries like Pymetis to intelligently split the network topology. We also plan to incorporate more sophisticated aggregation protocols. This includes implementing decentralized, peer-to-peer (P2P) aggregation to remove the central server and integrating secure aggregation techniques based on secure multi-party computation (SMPC) or homomorphic encryption to enhance privacy. Looking further ahead, we plan to integrate the testbed with container orchestration platforms like Kubernetes and introduce energy consumption models to enable the design of energy-efficient FL algorithms, a crucial requirement for battery-powered edge devices.

\bibliographystyle{IEEEtran}  
\footnotesize
\bibliography{ref}

\end{document}